\documentclass[aps,prd,reprint,twocolumn,superscriptaddress,showpacs]{revtex4-1}
\usepackage{amsfonts}
\usepackage{mathrsfs}
\usepackage{amsmath}
\usepackage{color}
\usepackage{natbib}
\usepackage{textcomp}
\usepackage{graphicx}
\usepackage{bm}
\usepackage{amssymb}

\usepackage{xspace}
\usepackage{epstopdf}
\usepackage{dcolumn}
\usepackage{longtable}
\usepackage{multirow}
\usepackage[colorlinks=true, letterpaper=true, pdfstartview=FitV, linkcolor=blue, citecolor=blue, urlcolor=blue]{hyperref}
\usepackage{float}

\makeatletter

\newcommand{\Rmnum}[1]{\expandafter\@slowromancap\romannumeral #1@}
\makeatother

\begin{document}

\title{Topological gimbal phonons in T-carbon}

 \author{Jing-Yang You}
\affiliation{Kavli Institute for Theoretical Sciences, and CAS Center for Excellence in Topological Quantum Computation, University of Chinese Academy of Sciences, Beijing 100190, China}
\affiliation{Department of Physics, Faculty of Science, National University of Singapore, 117551, Singapore}

\author {Xian-Lei Sheng}
 \email{xlsheng@buaa.edu.cn}
 \affiliation{School of Physics, Beihang University, Beijing 100191, China}
 
\author{Gang Su}
\email{gsu@ucas.ac.cn}
\affiliation{Kavli Institute for Theoretical Sciences, and CAS Center for Excellence in Topological Quantum Computation, University of Chinese Academy of Sciences, Beijing 100190, China}

\begin{abstract}
The topological metal states in electronic systems have been extensively studied, but topological phonons were explored only in few examples so far. Here, we expose for the first time that the topological nodal gimbal phonons, type-I and type-II Weyl phonons are simultaneously present in T-carbon, a recently realized new allotrope of carbon. At about 15.2 THz, we find that there exist three mutually intersecting nodal loops (named as nodal gimbal phonons) around $\Gamma$ point, and two pairs of type-I Weyl phonons on the boundary of Brillouin zone around each $X$ point. In addition, there exist three pairs of type-II Weyl phonons at about 14.5 THz around  each $L$ point. It is shown that these exotic topological phonons are protected by corresponding symmetries, and lead to topologically nontrivial surface states. Our findings not only afford plenty of intriguing topological phonon states in a simple material like T-carbon but also provide a new platform to study novel properties of topological phonons, which would facilitate further both experimental and theoretical works in future.
\end{abstract}
\maketitle


{\color{blue}\em Introduction}---Topological metals with symmetry protected band crossings have attracted much attention in both condensed matter physics and materials science~\cite{RMP_Das,RMP_Chiu,Burkov:2016wg,Yan2017review,Armitage2018}. The study is based on the analogy  between elementary particles in the relativity quantum field theory and low-energy emergent fermions in condensed matter.
With this analogy, Weyl and Dirac semimetals were discovered~\cite{Wan2011,Burkov2011,YoungSM2012,WangZJ2012,WangZJ2013,Liu2014,Weng2015,Lv2015}, 
which have twofold and fourfold degenerate nodal points, respectively, and around these points, low-energy electrons are similar to Weyl and Dirac fermions and exhibit fascinating physical effects similar to their counterparts in high-energy physics.
Depending on the dispersion of two crossing bands, the Weyl and Dirac semimetals can be classified into type-I or type-II~\cite{ZhangCW2015,Soluyanov:2015tc,LiS2017,ZhangXM2018}. 
In light of the dimension of the degeneracy manifold, the band crossings may exhibit zero-dimensional nodal point, 1D nodal lines~\cite{Weng2015a,Mullen2015,Kim2015,Fang2016} or even 2D nodal surfaces~\cite{Liang2016,Zhong2016,Wu2018,Gao2019}.

So far, topological band theory is mainly discussed in the context of electronic systems. Recently, topological phonons have also attracted much attention~\cite{Zhang2010,Li2012,Liu2017,Ji2017,Zhang2018,Jin2018,Xia2019,Liu2019,Zheng2020,Wang2020} owing to their potential applications in electron-phonon coupling, dynamic instability~\cite{Prodan2009}, and phonon diodes~\cite{Liu2017}. However, it is not easy to find a realistic material featuring topological phonons, because it requires the topological surface states of phonons that should be well separated from the bulk phonon spectrum. Carbon material may be a family of such ideal candidates due to their stable structures and excellent electronic and phonon properties. Nonetheless, the studies on topological phonons in carbon materials are still sparse.

T-carbon as a new carbon allotrope has been proposed theoretically in 2011~\cite{Sheng2011} and then successfully synthesized in experiments recently~\cite{Zhang2017,Xu2020}. Due to its unique structure, T-carbon was shown to possess versatile potential applications such as hydrogen storage~\cite{Sheng2011}, solar cells~\cite{Sun2019}, lithium ion batteries~\cite{Qin2019}, thermoelectrics~\cite{Yue2017,Qin2019}, photocatalyst~\cite{Ren2019,Alborznia2019}, seawater desalination~\cite{Zhou2020}, superconductivity~\cite{You2020}, etc.

In this work, we show that T-carbon exhibits exotic topological phonon states.
At about 15 THz, both nodal loops and type-I Weyl points coexist. Around $\Gamma$ point, there are three intersecting nodal loops. As the structure of the three nodal loops looks like a gimbal, it is thus coined as nodal gimbal phonons. Moreover, around the center ($X$) of the square at the boundary of Brillouin zone (BZ) there are two pairs of Weyl points, and because there are three independent X points in BZ, we have six pairs of such Weyl points in total. In addition, at 14.5 THz, three pairs of type-II Weyl points appear around the center ($L$) of hexagonal surface at the boundary of BZ. There are four independent $L$ points in BZ, leaving 12 pairs of type-II Weyl points. In terms of the lattice symmetries and $k\cdot p$ model analyses, these topological phonons are found to be protected by the corresponding symmetries and are materials-independent, which can thus be applicable to diamond because T-carbon has the same space group as the latter. Our results not only offer opportunities to study novel topological phonons in carbon materials, but also provide a new platform to explore the emergent physics due to the interplay between topological phonons and electron-phonon coupling and thermal transport.

{\color{blue}\em Calculation method}---Our first-principles calculations were based on density functional theory (DFT) as implemented in the Vienna \textit{ab initio} Simulation Package (VASP)~\cite{Kresse1996}, using the projector augmented-wave method~\cite{Bloechl1994}. The generalized gradient approximation with the Perdew-Burke-Ernzerhof~\cite{Perdew1996} realization was adopted for the exchange-correlation functional. The plane-wave cutoff energy was set to 550 eV. A Monkhorst-Pack k-point mesh~\cite{Monkhorst1976} with a size of 15$\times$15$\times$15 was used for the Brillouin zone (BZ) sampling. The crystal structure was optimized until the forces on the ions were less than 0.0001 eV/\AA. The phonon spectra were obtained with the PHONOPY package~\cite{Togo2015}, where a 2$\times$2$\times$2 supercell and a displacement of 0.01 \AA \ from the equilibrium atomic positions are used. The surface spectrum was calculated by using the Wannier functions and the iterative Green's function method~\cite{Marzari1997, Souza2001, Wu2018a, Sancho1985}.

\begin{figure}[!htbp]
  \centering
  \includegraphics[scale=0.45,angle=0]{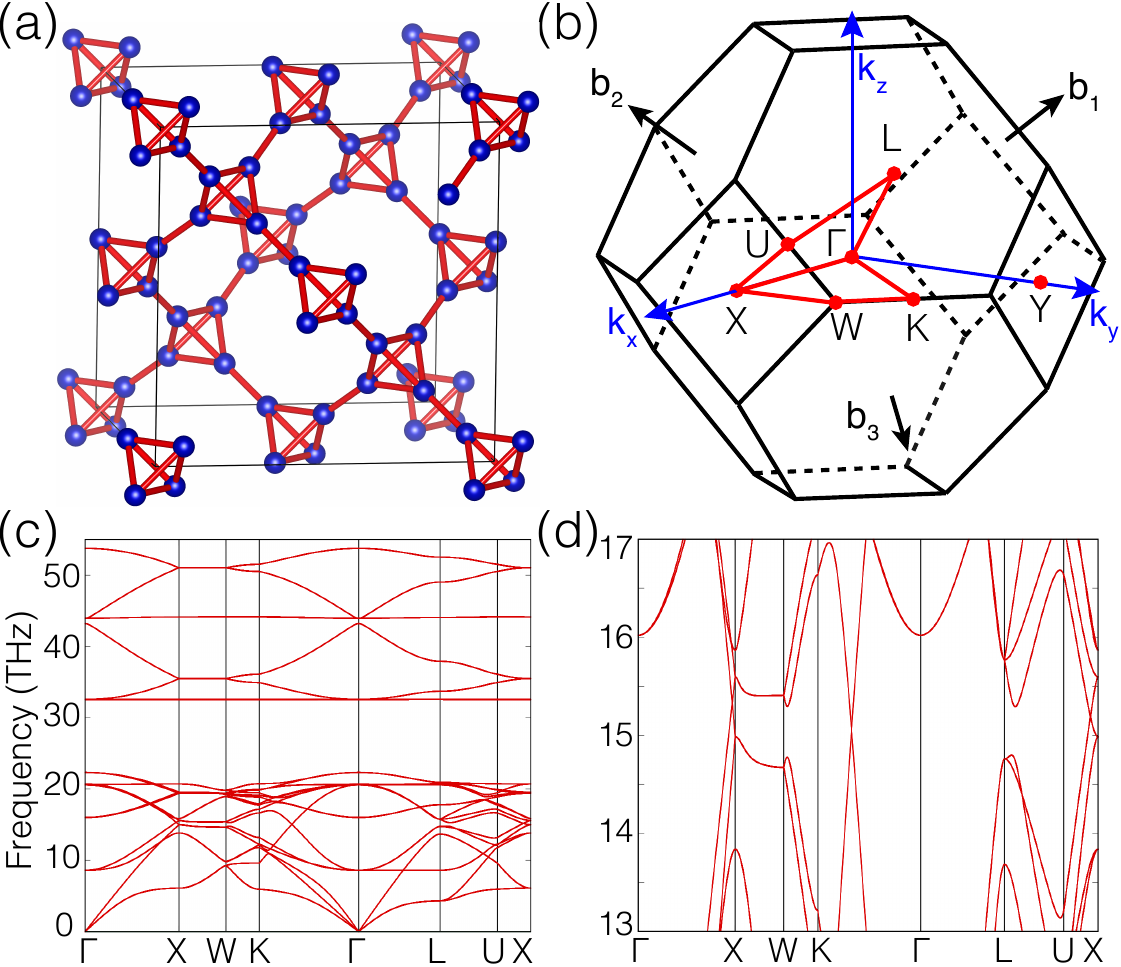}\\
  \caption{(a) The cubic crystalline structure of T-carbon. (b) The BZ of the primitive cell of T-carbon with high symmetry points and paths indicated. (c) The phonon spectra of T-carbon along high-symmetry path. (d) The enlarged views of phonon spectra of T-carbon in the range of 13 to 17 THz.}\label{fig1}
\end{figure}

{\color{blue}\em Phonon spectrum}---T-carbon possesses face-centered cubic lattice with the space group of $Fd\bar{3}m$ (No.227) corresponding to the point group $O_h$ as shown in Fig.~\ref{fig1}(a), which is a direct-gap semiconductor. The structure of T-carbon can be viewed as replacing each atom of cubic diamond with a tetrahedron formed by four carbon atoms, such that each primitive cell of T-carbon contains two tetrahedrons with eight carbon atoms. The optimized lattice constant $a$ is about 7.52 \AA. The three unit vectors are $\vec{a}=(l/2)(0,1,1), \vec{b}=(l/2)(1,0,1)$, and $\vec{c}=(l/2)(1,1,0)$, and the carbon atoms occupy the Wyckoff position $32e(x; x; x)$ with $x\sim0.0706$.

\begin{figure}[!htbp]
  \centering
  \includegraphics[scale=0.3,angle=0]{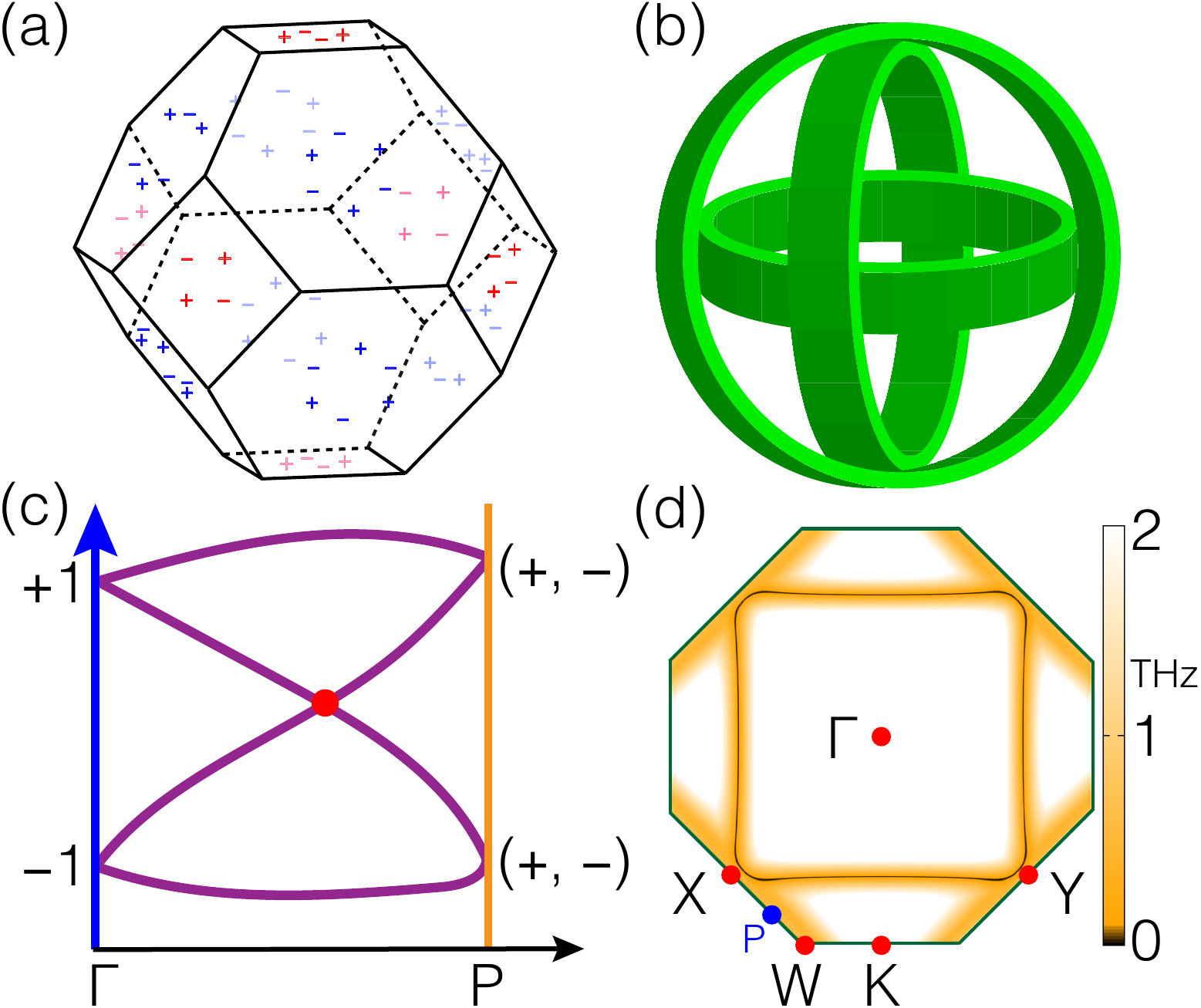}\\
  \caption{(a) The distribution of Weyl points at the boundaries of BZ in T-carbon, i.e, the $k_{x,y,z}=2\pi/a$ and $(\pm1,\pm1,\pm1)$ planes, where the red- and blue-colored signs represent the Weyl point at the square and hexagonal surfaces, respectively, and the "+" and "-" indicate the chirality of Weyl points. (b) Schematic depiction of the gimbal. (c) Schematic figure showing the band crossing along the paths connecting $\Gamma$ to some arbitrary point $P$ on path $X$-$W$. (d) Shape of the hourglass Weyl loop (black-colored loop) in $k_z=0$ plane obtained from the DFT calculations. The color map corresponds to the frequency difference between the two crossing phonon bands.}\label{fig2}
\end{figure}

The phonon spectra of T-carbon along high-symmetry paths [Fig.~\ref{fig1}(b)] of the BZ are plotted in Fig.~\ref{fig1}(c). It is clear that there is a large phonon band gap of about 10 THz between about 22 to 32 THz, and two flatbands exist at about 32 THz and 44 THz, which are related to the geometry of T-carbon that can be seen from its view of [110] direction~\cite{You2019}. As the contribution of high frequency phonons to the electron-phonon coupling and thermal transport is very low~\cite{You2020}, we will focus on the relative low-frequency phonons below 20 THz. From the enlarged view of phonon spectra within 13 THz and 17 THz as shown in Fig.~\ref{fig1}(d), we find that two phonon bands from optical branch cross linearly at about 15 THz in paths $\Gamma-X$, $K-\Gamma$ and $U-X$. After a careful inspection within $k_z=0$ plane, we observe that the crossing points are not isolated, but form a nodal loop in $k_z=0$ plane, which is with respect to the mirror reflection symmetry $\mathcal{M}_z$. The two crossing phonon bands within the mirror invariant plane $\mathcal{M}_z$ have opposite mirror eigenvalues $\pm1$. Because of the three-fold rotation symmetry along the [111] direction $C_3^{[111]}$, there should be another two nodal loops locating in $k_x$ = 0 and $k_y$ = 0 planes, respectively, which are also verified by our DFT calculations. These three rings are perpendicular to each other and intersect at six points on the coordinate axis, thus forming a gimbal as shown in Fig.~\ref{fig2}(b). The band crossing point in path $U-X$ will lead to another three crossing points related to the four-fold rotation symmetry along $x$ axis ($C_{4x}$) in $k_x=2\pi/a$ plane. Considering other two four-fold rotation symmetries along $y$ and $z$ axis, we can obtain two pairs of Weyl nodes within each square of the boundary of BZ. Thus, there will be a total of six pairs of Weyl phonons in BZ as shown in Fig.~\ref{fig2}(a). The band crossing point at about 14.5 THz in path $L-U$ will lead to other six crossing points in the $(111)$ plane. There are four such planes in BZ, thus we can obtain  a total of twelve pairs of Weyl phonons in BZ [Fig.~\ref{fig2}(a)].

Besides the above band crossing points, it is noted that the double degeneracy appears on the diagonals of the squares at the boundary of BZ, such as $X$-$W$ in Fig.~\ref{fig1}. One may see that any $k$ point on this path is invariant under both glide mirror symmetries $\tilde{\mathcal{M}}_x$ and $\tilde{\mathcal{M}}_z$. The commutation relation between $\tilde{\mathcal{M}}_x$ and $\tilde{\mathcal{M}}_z$ is given by
\begin{equation}
\tilde{\mathcal{M}}_x\tilde{\mathcal{M}}_z=T_{\overline{\frac{1}{2}},0,\frac{1}{2}}\tilde{\mathcal{M}}_z\tilde{\mathcal{M}}_x
\end{equation}
where $T_{\overline{\frac{1}{2}},0,\frac{1}{2}}=e^{i(k_x-k_z)/2}$ represents the translation along the $[\overline{1}01]$ direction in half unit cell. Along $X$-$W$ and $k_z=0$, we have $k_x=2\pi/a$; hence, $T_{\overline{\frac{1}{2}},0,\frac{1}{2}}=-1$. Therefore, the two glide mirror symmetries anticommutate along the $X$-$W$ path. As a result, for any energy eigenstate $|u\rangle$  with $\tilde{\mathcal{M}}_z$ of eigenvalue $g_z$, it must have a degenerate partner $\tilde{\mathcal{M}}_x|u\rangle$ with $\tilde{\mathcal{M}}_z$ of eigenvalue $-g_z$. This proves that the double degeneracy on $X$-$W$ path is guaranteed by the symmetries.

{\color{blue}\em Gimbal phonons}---Around $\Gamma$ point, there are three intersecting nodal loops for T-carbon in the planes $k_x=0$, $k_y=0$ and $k_z=0$, respectively. 
In the following, we take the loop in $k_z=0$ plane as an example to show that the nodal loop on this plane is protected by symmetries and is caused by band inversion. Each $k$ point in the $k_z=0$ plane is invariant under $\tilde{\mathcal{M}}_z: (x, y, z) \rightarrow (x+\frac{1}{4}, y+\frac{1}{4}, -z+\frac{1}{4})$, so any Bloch state $|u\rangle$ at momentum $k$ can be chosen as the eigenstate of $\tilde{\mathcal{M}}_z$. One finds that $(\tilde{\mathcal{M}}_z)^2=T_{\frac{1}{2}\frac{1}{2}0}=e^{-i\cdot\frac{1}{2}(k_x+k_y)a}$. Hence, the eigenvalues of $\tilde{\mathcal{M}}_z$ are given by $g_z=\pm e^{-i\cdot\frac{1}{4}(k_x+k_y)a}$. There are four time reversal invariant momenta (TRIM) points in the $k_z=0$ plane, labeled as $\Gamma$, $X(k_x=2\pi/a)$, $Y(k_y=2\pi/a)$ and $L$ in Fig.~\ref{fig1}(b). At these points, the bands must form degenerate Kramers pairs due to the presence of $\mathcal{T}$. Let us consider the $\tilde{\mathcal{M}}_z$ eigenvalues $g_z$ at these points. For example, at $X(Y)$, we have $g_z = \pm i$, so each Kramers pair $|u\rangle$ and $\mathcal{T}|u\rangle$ must have opposite $g_z$. However, at $\Gamma$ point, since $g_z = \pm 1$, each Kramers pair $|u\rangle$ and $\mathcal{T}|u\rangle$ must share the same $g_z$. Due to this different pairing at $\Gamma$ and $X(Y)$, there must be a switch of partners between two pairs when going from $\Gamma$ to $X(Y)$, during which the four bands must be entangled to form the hourglass dispersion. However, for one path $\ell$ connecting $\Gamma$ to some arbitrary point $P$ on the $X$-$W$ path, each state at $P$ has a double degeneracy with opposite $\tilde{\mathcal{M}}_z$ eigenvalues $\pm g_z$, which is labeled as $(+, -)$ in Fig.~\ref{fig2}(c). Due to the degeneracy with the same $\tilde{\mathcal{M}}_z$ eigenvalues at $\Gamma$, the switch of partners between two pairs on path $\ell$ guarantees the hourglass dispersion. On the other hand, the corresponding four states are not required to be degenerate on path $W$-$K$, where the $\tilde{\mathcal{M}}_z$ eigenvalues are $(+, +, -, -)$ for the states in descending order. Focusing on the middle two bands, they have opposite  $\tilde{\mathcal{M}}_z$ eigenvalues with inverted ordering between $\Gamma$ and some arbitrary point $A$ on path $X$-$W$, and as a result, they must cross. Thus, the crossing point will trace out a hybrid nodal loop on the $k_z=0$ centered at $\Gamma$, where the hybrid nodal loop contains Weyl and hourglass Weyl phonons as shown in Fig.~\ref{fig2}(b). Similarly, other two hybrid nodal loops appearing on the $k_x=0$ and $k_y=0$ planes are also symmetry-protected. 

\begin{figure}[!htbp]
  \centering
  \includegraphics[scale=0.42,angle=0]{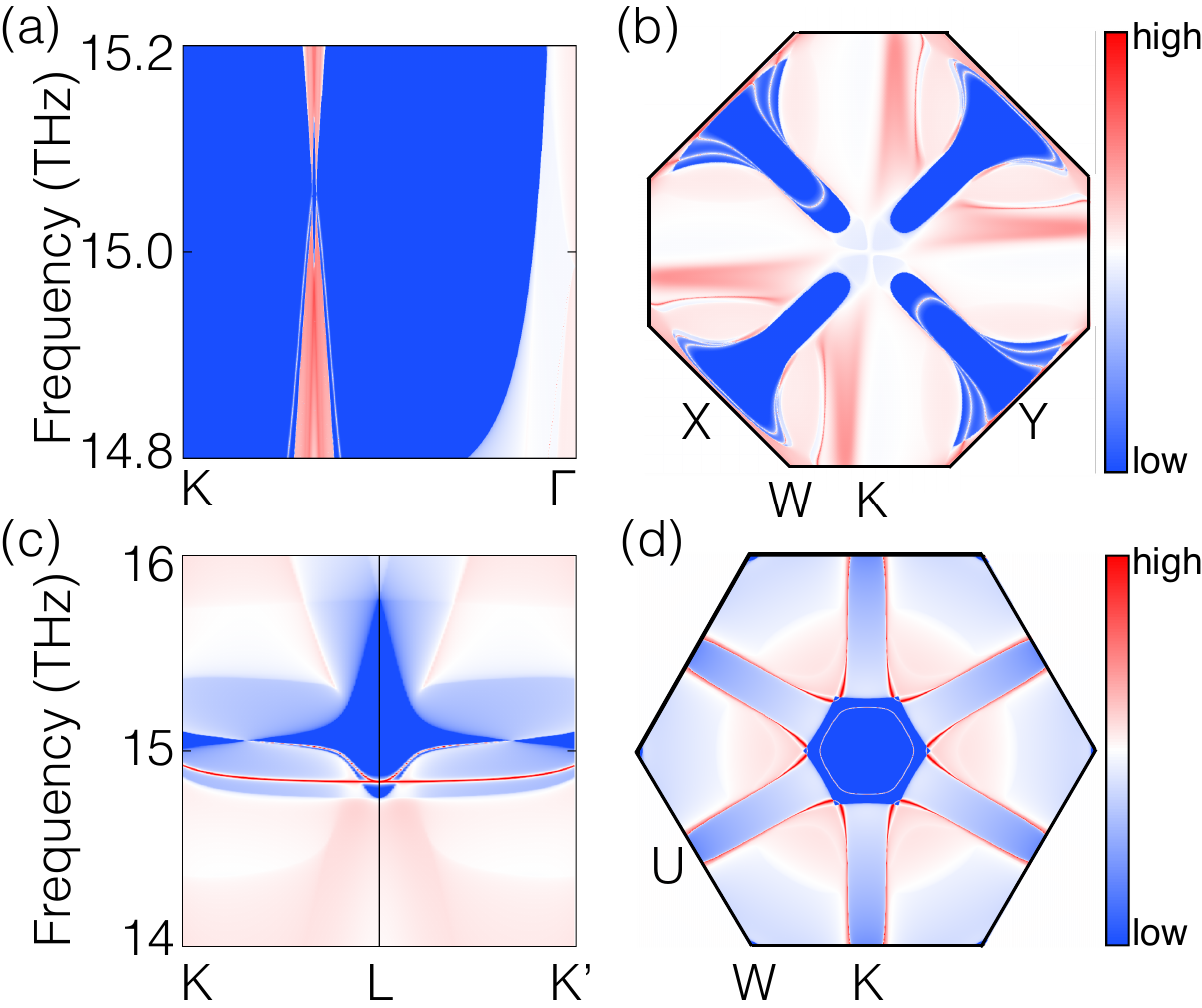}\\
  \caption{(a) Local density of states (LDOS) and (b) a constant energy slice at 15 THz projected on the semi-infinite (001) surface of T-carbon. (c) LDOS and (d) a constant energy slice at 15 THz projected on the semi-infinite (111) surface.}\label{fig3}
\end{figure}

Figure~\ref{fig2}(d) shows the shape of the nodal loop obtained from DFT in $k_z=0$ plane within BZ. The surface of a nodal line semimetal features the drumhead like states. In Figs.~\ref{fig3}(a) and \ref{fig3}(c), we show the phonon surface states of T-carbon on (001) and (111) surfaces, respectively. Indeed, one observes the drumhead surface bands that emanate from the bulk nodal points, which connects the two nodal lines through the surface BZ boundary. In Figs.~\ref{fig3}(b) and \ref{fig3}(d), we plot the constant frequency slice at 15 THz, which cuts through the drumhead, forming a few arcs or circles, because the drumhead is not completely flat in frequency. 

{\color{blue}\em Symmetry protection}---The nodal gimbal is protected by time reversal symmetry and mirror symmetries $\mathcal{M}_x$, $\mathcal{M}_y$ and $\mathcal{M}_z$. To see this more clearly, it is better to write down a $k \cdot p$ low-energy Hamiltonian around $\Gamma$ point. The symmetry at $\Gamma$ point is characterized by $O_{h}$ point group, whose generators consist of three mirror planes $\mathcal{M}_x: (x, y, z) \to (-x, y, z)$, $\mathcal{M}_y:(x, y, z) \to (x, -y, z)$, and $\mathcal{M}_z:(x, y, z) \to (x, y, -z)$ and three $C_{4}$ operators. We can construct a minimal low-energy model for the two crossing bands around $\Gamma$:
\begin{equation}
\mathcal{H}_\Gamma(\mathbf{k})=\varepsilon_0(\mathbf{k})+\sum_{i = x, y, z}d_i(\mathbf{k})\sigma_i,
\end{equation}
where $d_i$($\mathbf{k}$) ($i=x,y,z$) are real functions of momentum $\mathbf{k}$ and the vector $\mathbf{k}$ is measured relative to the $\Gamma$ point. The first term is proportional to the identity matrix with a real function $\varepsilon_0(\mathbf{k})$. For such a phonon system, the time-reversal symmetry operator is represented by $\mathcal{T}=\mathcal{K}$ which is the complex conjugate satisfying $\mathcal{T}^2=1$. With the above constraints, the Hamiltonian should satisfy the following requirements:
\begin{equation}\label{eqT}
\mathcal{T}\mathcal{H}_\Gamma(\mathbf{k})\mathcal{T}^{-1}= \mathcal{H}_\Gamma(-\mathbf{k}),
\end{equation}

\begin{equation}\label{eqMx}
\mathcal{M}_x\mathcal{H}_\Gamma(\mathbf{k})\mathcal{M}_x^{-1}= \mathcal{H}_\Gamma(-k_x, k_y, k_z),
\end{equation}

\begin{equation}\label{eqMy}
\mathcal{M}_y\mathcal{H}_\Gamma(\mathbf{k})\mathcal{M}_y^{-1}= \mathcal{H}_\Gamma(k_x, -k_y, k_z),
\end{equation}

\begin{equation}\label{eqMz}
\mathcal{M}_z\mathcal{H}_\Gamma(\mathbf{k})\mathcal{M}_z^{-1}= \mathcal{H}_\Gamma(k_x, k_y, -k_z).
\end{equation}

Equation~(\ref{eqT}) requires that $d_y(\mathbf{k})$ is an odd function of $\mathbf{k}$, while $d_{x,z}(\mathbf{k})$ are  even functions of $\mathbf{k}$. The eigenfunctions of the two crossing bands are also eigenfunctions of mirror symmetries $\mathcal{M}_x$, $\mathcal{M}_y$ and $\mathcal{M}_z$. The first-principles calculations show that the irreducible representations of the two crossing bands are opposite. Thus, the matrix representation of the three mirror operators could be $\sigma_z$.  Up to the third order, the Hamiltonian reads
\begin{equation}\label{eqH}
\mathcal{H}_\Gamma(\mathbf{k})=\varepsilon_0(\mathbf{k})+\\
\left(
\begin{array}{cc}
d_z(\mathbf{k}) & -ibk_xk_yk_z \\
ibk_xk_yk_z  &  -d_z(\mathbf{k})
\end{array}
\right)\\
,
\end{equation}
where $\varepsilon_0(\mathbf{k})=a_0+a_1(k_x^2+k_y^2+k_z^2)$, $d_z(\mathbf{k})=c_0+c_1(k_x^2+k_y^2+k_z^2)$. The parameters $a_i$, $c_i$ ($i$=0,1) and $b$ can be
derived by fitting the dispersions to those of first-principles calculations.
The two bands around $\Gamma$ point  with the inverted structure lead to $c_0 > 0$ and $c_1 < 0$, which is essential for the existence of nodal loops. On the plane $k_x=0$, Eq.~(\ref{eqH}) leads to 
\begin{equation}
c_0+c_1(k_y^2+k_z^2)=0,
\end{equation}
which gives the band-crossing points to form a circle in the $k_y$-$k_z$ plane. Similarly, on the plane $k_y=0$ and $k_z=0$, from Eq.~(\ref{eqH}) one gets
\begin{equation}
c_0+c_1(k_x^2+k_z^2)=0,
\end{equation}

\begin{equation}
c_0+c_1(k_x^2+k_y^2)=0,
\end{equation}
which leads to other two nodal loops in the $k_y=0$ and $k_z=0$ planes. 

Based on the $k\cdot p$ model, it can be revealed that there is no other band crossing points around $\Gamma$ except the nodal gimbal on the three mirror planes. In general, the eigenvalues of Eq.~(\ref{eqH}) take the form
\begin{equation}
  E=\varepsilon_0(\mathbf{k}) \pm \sqrt{(d_z(\mathbf{k}))^2 + (bk_xk_yk_z)^2}.
\end{equation}
To get band crossing points, both terms $(d_z(\mathbf{k}))^2$ and $(bk_xk_yk_z)^2$ should be zero. Then, the second term requires either $k_x=0$ or $k_y=0$ or $k_z=0$. Thus, the nodal points only exist on the three planes. Since the k-points here are measured from $\Gamma$, the three k-planes correspond to $k_x=0$, $k_y=0$ and $k_z=0$. These discussions have been confirmed by the first-principles calculations [Fig.~\ref{fig2}].

\begin{figure*}[!!!htbp]
  \centering
  \includegraphics[scale=0.5,angle=0]{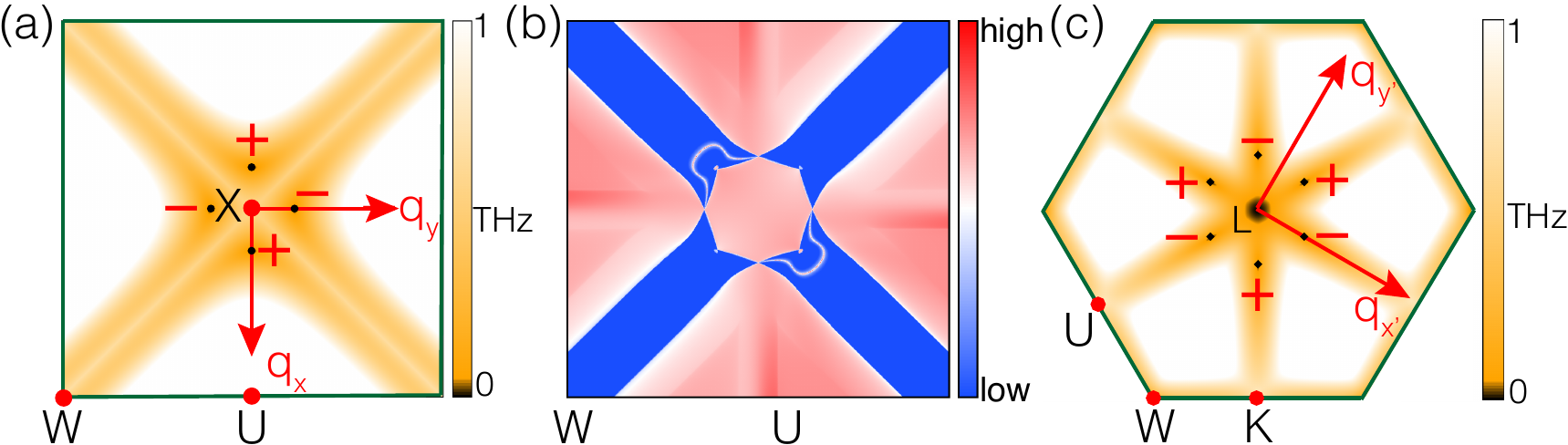}\\
  \caption{(a) Distribution of Weyl phonons in $k_x=2\pi/a$ plane of T-carbon obtained from the DFT calculations, where the Weyl phonons with opposite chirality are marked as "+" and "-", respectively, and the redefined Cartesian coordinate system $(q_x, q_y, q_z)$ in $k_x=2\pi/a$ plane is also indicated. The color map corresponds to the energy difference between the two crossing bands. (b) The phonon surface arcs projected on the semi-infinite (100) surface of T-carbon at a constant frequency slice 15.2 THz. (c) Distribution of Weyl phonons at about 14.5 THz in (111) plane obtained from the DFT calculations, where the Weyl phonons with opposite chirality are marked as "+" and "-", respectively, and the redefined Cartesian coordinate system $(q_{x^{\prime}}, q_{y^{\prime}}, q_{z^{\prime}})$ in (111) plane is indicated. The color map corresponds to the energy difference between the two crossing bands.}\label{fig4}
\end{figure*}

{\color{blue}\em Weyl phonons}---From Fig.~\ref{fig1}(d), one observes that besides the three nodal loops (gimbal phonons) around $\Gamma$, there is other band crossing point along $U$-$X$ path at about 15.2 THz. From first-principles calculations, it is found that the two phonon bands possess different irreducible representations of $C_2$ rotation symmetry preserved by $U$-$X$. Therefore, it is a nontrivial band crossing, leading to a nodal Weyl point. Since the fourfold rotation symmetry along $\Gamma$-$X$, there should exist another three Weyl points in the $q_x=2\pi/a$ plane, which has been confirmed by our first-principles calculations. The distribution of the four Weyl points is shown in Fig.~\ref{fig4}(a) with the chirality $\mathcal{C}$ = +1 or $-1$ marked as ``+" or ``$-$". One can see that there are two Weyl points of  +1 and two of $-1$.  The phonon surface arcs projected on the semi-infinite (100) surface at a constant frequency slice 15.2 THz are plotted in Fig.~\ref{fig4}(b). It is clear that there are two Fermi arcs connected with two opposite Weyl points at the ends, which is the fingerprint of nontrivial Weyl phonons. 

In the following, we show that the Weyl phonons in the squares at the boundary of BZ are protected by symmetries. We take the Weyl phonons in $k_x=2\pi/a$ for an example. The two crossing branches of phonons generally can be described by a 2$\times$2 $k\cdot p$ Hamiltonian,
\begin{eqnarray}
\mathcal{H}_X(\mathbf{q}) = \sum_{i = x, y, z}f_i(\mathbf{q})\sigma_i,
\end{eqnarray}
where $f_i(\mathbf{q})$ are real functions, and $\sigma_i$ are Pauli matrices. Note that $q_x$, $q_y$ and $q_z$ are measured relative to $X$,  with $q_x$ along the $X$-$U$ direction and $q_z$ along the $k_x$ direction as shown in Fig.~\ref{fig4}(a).

We first consider the two-fold rotation symmetry along $q_z(k_x)$ axis $C_{2z}(C_{2x})$. As the two crossing branches belong to opposite eigenvalues of $C_2$,  the $C_{2z}$ can be chosen as $\sigma_z$. 
The $q_z=2\pi/a(k_x=2\pi/a)$ plane is invariant under the combination of $C_{2z}$ and $\mathcal{T}$, and $C_{2z}\mathcal{T}$ can be represented by $\sigma_z\mathcal{K}$, where $\mathcal{K}$ is the complex conjugate operator. The symmetry $C_{2z}\mathcal{T}$ requires 
\begin{eqnarray}
C_{2z}\mathcal{T}\mathcal{H}_X(\mathbf{q})(C_{2z}\mathcal{T})^{-1}=\mathcal{H}_X(C_{2z}\mathcal{T}\mathbf{q}),
\end{eqnarray}
which leads to 
\begin{eqnarray}
-f_x(q_x, q_y, q_z)=f_x(q_x, q_y, -q_z),\\
f_{y,z}(q_x, q_y, q_z)=f_{y,z}(q_x, q_y, -q_z).
\end{eqnarray}
For the $q_z=2\pi/a$ plane, we have $q_z=-q_z$ because of the periodic condition of $\mathcal{H}_X(\mathbf{q})$. In this case, $f_x(q_x, q_y, 2\pi/a)\equiv 0$. Besides, the path $X$-$U$ is invariant under $C_{2x}$ and $C_{2y}\mathcal{T}$, which constrains $\mathcal{H}_X(q_x, q_y, 2\pi/a)$ and gives rise to
\begin{eqnarray}
-f_{y,z}(q_x, q_y, 2\pi/a)=f_{y,z}(q_x, -q_y, 2\pi/a)\label{eqCx},\\
-f_{y,z}(q_x, q_y, 2\pi/a)=f_{y,z}(-q_x, q_y, 2\pi/a)\label{eqCy}.
\end{eqnarray}
It turns out that Eqs.~(\ref{eqCx}) and (\ref{eqCy}) ensure the crossing points to be along the $q_x$ or $q_y$ axis, which are protected by the symmetries $C_{2x}$, $C_{2y}$ and $C_{2z}$ and $\mathcal{T}$. Thus, there exist two pair of Weyl points around $X$. Through a similar analysis, the combination of $C_2$ and $\mathcal{T}$ can also dictate that the Weyl phonons can locate along the high-symmetry lines $X$-$U$ in the $k_y=2\pi/a$ and $k_z=2\pi/a$ planes.

{\color{blue}\em Type-II Weyl phonons in (111) plane}---We take the (111) plane as an example to reveal the type-II Weyl phonons in T-carbon at about 14.5 THz, and the distribution is shown in Fig.~\ref{fig4}(c) with the chirality $\mathcal{C}$ = +1 or $-1$ marked as ``+" or ``$-$". 

For simplicity, we redefine a Cartesian coordinate system $(q_{x^{\prime}}, q_{y^{\prime}}, q_{z^{\prime}})$ in (111) plane as shown in Fig.~\ref{fig4}(c).
The two crossing branches of phonons can be generally described by a 2$\times$2 $k\cdot p$ Hamiltonian
\begin{eqnarray}
\mathcal{H}_L(\mathbf{q^{\prime}}) = \sum_{i = x, y, z}m_i(\mathbf{q^{\prime}})\sigma_i,
\end{eqnarray}
where $m_i(\mathbf{q^{\prime}})$ are real functions, and $\sigma_i$ are Pauli matrices.

Let us first consider the mirror symmetry $\mathcal{M}_{y^{\prime}}$. As the two crossing branches belong to opposite eigenvalues of $\mathcal{M}_{y^{\prime}}$, $\mathcal{M}_{y^{\prime}}$ can be chosen as $\sigma_z$. The $q_{x^{\prime}}$ axis is invariant under $\mathcal{M}_{y^{\prime}}$. The symmetry $\mathcal{M}_{y^{\prime}}$ requires 
\begin{eqnarray}
\mathcal{M}_{y^{\prime}}\mathcal{H}_L(\mathbf{q^{\prime}})(\mathcal{M}_{y^{\prime}})^{-1}=\mathcal{H}_L(q_{x^{\prime}}, -q_{y^{\prime}}, q_{z^{\prime}}),
\end{eqnarray}
which leads to 
\begin{eqnarray}
-m_{x,y}(q_{x^{\prime}}, q_{y^{\prime}}, q_{z^{\prime}})=m_{x,y}(q_{x^{\prime}}, -q_{y^{\prime}}, q_{z^{\prime}}),\\
m_z(q_{x^{\prime}}, q_{y^{\prime}}, q_{z^{\prime}})=m_z(q_{x^{\prime}}, -q_{y^{\prime}}, q_{z^{\prime}}).
\end{eqnarray}
On the other hand, the $q_{x^{\prime}}$ axis is invariant under the combination of $C_{2y^{\prime}}$ and $\mathcal{T}$, which further constrains $\mathcal{H}_L(q_{x^{\prime}}, q_{y^{\prime}}, q_{z^{\prime}})$ and leads to
\begin{eqnarray}
-m_z(q_{x^{\prime}}, q_{y^{\prime}}, q_{z^{\prime}})=m_z(q_{x^{\prime}}, -q_{y^{\prime}}, q_{z^{\prime}}).
\end{eqnarray}
Consequently, we have $m_z\equiv 0$ in the $q_{x^{\prime}}$ axis. Besides, for the $q_{x^{\prime}}$ axis, $q_{y^{\prime}}=0$, thus $m_{x,y} = 0$.
Therefore, the crossing points should exist along the $q_{x^{\prime}}$ axis, which are protected by the symmetries $\mathcal{M}_{y^{\prime}}$, $C_{2y^{\prime}}$ and $\mathcal{T}$. In terms of the three-fold symmetry along the $q_{z^{\prime}}$ axis $C_{3z^{\prime}}$, the other two pairs of type-II Weyl phonons can also be obtained.



{\color{blue}\em Summary}---We have studied the topological states of phonons in carbon materials for the first time, and taken T-carbon as an instance to explore its exotic topological phonon states in detail. It is surprising that topological gimbal, type-I and type-II Weyl phonons exist simultaneously in T-carbon. At about 15.2 THz, T-carbon has both nodal loops and Weyl points. Around the $\Gamma$ point, there are three intersecting nodal loops, named as nodal gimbal. On the square at the boundary of BZ, there are two pairs of Weyl points around each $X$ point. We found that three independent $X$ points appear in the first BZ, and thus six pairs of Weyl points exist in total. In addition, at about 14.5 THz, there are three pairs of type-II Weyl points on each hexagon at the boundary of BZ. Since four independent $L$ points are present in the first BZ, there are 12 pairs of type-II Weyl points totally. Importantly, these nodal loops and Weyl points are protected by corresponding symmetries. In light of our analyses, the topological phonon properties observed in T-carbon can be extended to other materials with the same symmetry such as cubic diamond. This present study not only reveals rich topological phonon states in a single material like T-carbon but also provides a new platform to tackle exotic topological phonons in simple condensed matter systems, which would spur further experimental and theoretical works in future. 

{\color{blue}\em Acknowledgement}---This work is supported in part by the National Key R\&D Program of China (Grant No. 2018YFA0305800), the Strategic Priority Research Program of the Chinese Academy of Sciences (Grants No. XDB28000000), the National Natural Science Foundation of China (Grant No.11834014), and Beijing Municipal Science and Technology Commission (Grant No.  Z191100007219013). 

%

\end{document}